\documentstyle[aps,12pt]{revtex}
\begin{document}
\title{{\bf Klein Tunnelling and The Klein Paradox }}
\author{{\bf A\ Calogeracos}}
\author{(acal@hol.gr)}
\author{NCA\ Research Consultants, PO Box 61147,Maroussi 151 22, Greece}
\author{{\bf N Dombey}}
\author{(normand@sussex.ac.uk)}
\author{Centre for Theoretical Physics, University of Sussex, UK}
\date{SUSX-TH-97-019}
\maketitle
\medskip\
\begin{abstract}
The Klein paradox is reassessed by considering the properties of a finite
square well or barrier in the Dirac equation. It is shown that spontaneous
positron emission occurs for a well if the potential is strong enough. The
vacuum charge and lifetime of the well are estimated. If the well is wide
enough, a seemingly constant current is emitted. These phenomena are
transient whereas the tunnelling first calculated by Klein is
time-independent. Klein tunnelling is a property of relativistic wave
equations, not necessarily connected to particle emission. The Coulomb
potential is investigated in this context: it is shown that a heavy nucleus
of sufficiently large $Z$ will bind positrons. Correspondingly, it is
expected that as $Z$ increases the Coulomb barrier will become increasingly
transparent to positrons. This is an example of Klein tunnelling.
\newline
\newline

\center{To be published in International Journal of Modern Physics A}
\end{abstract}

\newpage

\section{The Klein Paradox}

\noindent Klein \cite{klein} considered the reflection and transmission of
electrons of energy $E$ incident on the potential step $V(x)=V,x>0;%
\,V(x)=0,x<0$ for $E<V$ in the one-dimensional time-independent Dirac
equation which can be written \cite{cdi} in terms of the usual Pauli
matrices as 
\begin{equation}
(\sigma _x\frac \partial {\partial x}-\sigma _z(E-V(x))+m)\psi (x)=0
\label{dir}
\end{equation}
He found that the reflection and transmission coefficients $R_S,T_S$ when
the step height $V>2m$ were given by 
\begin{equation}
R_S=%
%TCIMACRO{\QDOVERD( ) {1-\kappa }{1+\kappa } }
%BeginExpansion
{\displaystyle {1-\kappa  \overwithdelims() 1+\kappa }}
%EndExpansion
^2\qquad T_S=\frac{4\kappa }{(1+\kappa )^2}  \label{Rs/Ts}
\end{equation}
\noindent where $\kappa $ is the kinematic factor 
\begin{equation}
\kappa =\frac pk\frac{E+m}{E+m-V}=\sqrt{\frac{(V-E+m)}{(V-E-m)}\frac{(E+m)}{%
(E-m)}}  \label{kappa}
\end{equation}
\noindent $k=\sqrt{E^2-m^2}$ is the momentum of the incident electron and $%
p=-\sqrt{(V-E)^2-m^2}$ is the momentum of the transmitted particle for $x>0.$
Provided $p$ is negative\footnote{%
Pauli pointed out to Klein \cite{klein} the necessity of choosing $p$ to be
negative so that the group velocity of the particle for $x>0$ was positive.}%
, $R_S$ and $\,T_S$ are both positive and satisfy $\,R_S+T_S=1.$ Telegdi 
\cite{tel} has recently reviewed the history of the Klein paradox,
concentrating on the inaccuracies which followed from choosing the wrong
sign of $p.$ Nowadays the paradox is taken to lie in the result that as $%
V\rightarrow \infty $ for fixed $E,\,T_S$ tends to a non-zero limit. The
physical essence of the Klein paradox thus lies in the prediction that
fermions can pass through large repulsive potentials without exponential
damping. This we call Klein tunnelling.

\smallskip\ 

\noindent There have been various attempts over the last seventy years to
explain the Klein paradox. The preferred explanation given by Telegdi and in
textbooks \cite{grib} is due to Hansen and Ravndal \cite{hans}: the
left-right asymmetric step emits electron-positron pairs just as a constant
electric field does \cite{schwing} and it is the inability of an observer to
distinguish between this emission and the static scattering properties of
the step which gives rise to the paradox. $T_S$ measures the positrons
emitted by the step while $R_S$ is not unity because the Pauli principle
causes destructive interference between the emitted electron and the
reflected electron. In the absence of particle production $R_S\,$would
indeed be $1$ and $T_S$ would be $0.$

\smallskip\ 

\noindent These claims are investigated in this paper. We begin from a
viewpoint which sees no connection between left-right asymmetry and quantum
tunnelling, nor indeed any necessary connection between either of these and
particle production. We will seek to show that Klein tunnelling is
characteristic of relativistic wave equations as such, because negative
energy states really do represent physical particles. Our approach in this
respect is similar to that of Jensen et al \cite{jens}. We therefore
investigate a class of problems in which Klein tunnelling takes place even
in the absence of particle production. Klein was unfortunate in that he
considered a pathological problem which gives misleading results: the
potentials we consider can be very similar to a Klein step but are better
defined and both their time-independent and emission properties can be found
simply using Eq.(\ref{dir}) in the presence of a four-potential $V(x)$.
Particle emission from strong potentials is well-understood in terms of
supercritical positron production \cite{grib}. A full field theoretic
treatment is given in our earlier paper \cite{cdi} but provided we restrict
ourselves to potentials which are switched on adiabatically we manage here
with first quantisation (except in the Appendix where we derive the relation
between the transmission coefficient $T_S$ and the current emitted by the
step). Finally we look at Klein tunnelling in the context of the Coulomb
potential.

\section{Scattering by a Square Barrier}

\noindent Consider the square barrier $V(x)=V,|\,x\,|<a;V(x)=0,|\,x\,|>a$
instead of the Klein step. For $ma>>1$ this potential may be expected to
give similar results to those found by Klein. It does, but the results are
perhaps surprising.

\smallskip\ 

\noindent It is easy to show that the reflection and transmission
coefficients for the square barrier are given by \cite{jens} 
\begin{eqnarray}
R &=&\frac{(1-\kappa ^2)^2\sin ^2(2pa)}{4\kappa ^2+(1-\kappa ^2)^2\sin
^2(2pa)}\qquad  \label{barr} \\
T &=&\frac{4\kappa ^2}{4\kappa ^2+(1-\kappa ^2)^2\sin ^2(2pa)}
\end{eqnarray}

\noindent Note that Klein tunnelling is enhanced for a barrier: if

\begin{equation}
2pa=N\pi  \label{quant}
\end{equation}

\noindent corresponding to $E_N=V-\sqrt{m^2+N^2\pi ^2/4a^2}$ the electron
passes right through the barrier with no reflection: this is called a
transmission resonance \cite{cdi}. As $a$ becomes very large for fixed $m,E$
and $V$, we can average over the phase angle $pa$ to find the limit

\begin{equation}
R_\infty =\frac{(1-\kappa ^2)^2}{8\kappa ^2+(1-\kappa ^2)^2}\qquad T_\infty =%
\frac{8\kappa ^2}{8\kappa ^2+(1-\kappa ^2)^2}  \label{inf}
\end{equation}

\noindent It may seem unphysical that $R_\infty $ and $T_\infty $ are not
the same as $R_S$ and $T_S$ but it is not: it is well known in
electromagnetic wave theory \cite{stratton} that reflection off a
transparent barrier of large but finite width (with 2 sides) is different
from reflection off a transparent step (with 1 side). $R_\infty $ and $%
T_\infty $ thus involve Klein tunnelling but they arise from a more physical
problem than the Klein step. The zero of potential is properly defined for a
barrier whereas it is arbitrary for a step and the energy spectrum of a
barrier is easily calculable. Emission from a barrier or well is described
by supercriticality: the condition when the ground state energy of the
system overlaps with the continuum ($E=m$ for a barrier; $E=-m$ for a well)
and so any connection between particle emission and the time-independent
scattering coefficients $R$ and $T$ can be investigated.

\section{Fermionic Emission from a Square Well/Barrier}

\noindent We discussed this topic in our previous paper \cite{cdi} which we
refer to as CDI. We quickly review the argument. Spontaneous fermionic
emission is a non-static process and in the case of a seemingly static
potential, it is necessary to ask how the potential was switched on from
zero. We follow CDI in turning on the potential adiabatically.

\smallskip\ 

\noindent The bound state spectrum for the well $V(x)=-V,|\,x\,|<a;V(x)=0,|%
\,x\,|>a$ is easily obtained \cite{cdi}: there are even and odd solutions
given by the equations

\begin{equation}
\tan pa=\sqrt{\frac{(m-E)(E+V+m)}{(m+E)(E+V-m)}}  \label{even}
\end{equation}

\begin{equation}
\tan pa=-\sqrt{\frac{(m+E)(E+V+m)}{(m-E)(E+V-m)}}  \label{odd}
\end{equation}

\noindent where now the well momentum is given by $p^2=(E+V)_{}^2-m^2$. We
have changed the sign of $V$ so that it is now attractive to electrons
rather than positrons in order to conform with other authors who have
studied supercritical positron emission rather than electron emission .

\smallskip\ 

\subsection{Narrow Well}

\noindent The simplest case to discuss is a very narrow deep well $%
V(x)=-\lambda \delta (x)$ which is the limit of a square well with $\lambda
=2Va$. The bound states are then given for even $(e)$ and odd ($o)$ wave
functions by

\begin{equation}
E=m\cos \lambda \quad (e)\qquad E=-m\cos \lambda \quad (o)  \label{delta}
\end{equation}

\noindent When the potential is initially turned on and $\lambda $ is small
there is one bound state just below $E=m$. As $\lambda $ increases, $E$
decreases and at $\lambda =\pi /2$ $,E$ reaches zero. For $E>\pi /2$, $E$
becomes negative and if the well were originally vacant, we now have the
absence of a negative energy state which we must interpret as the presence
of a (bound) positron according to Dirac's hole theory. We can use the
anticommutation relations for the electron field \cite{cdi} to write the
charge $Q$ as $Q=Q_0+Q_p$ where $Q_0$ is a c-number called the vacuum charge 
\cite{stone}\cite{blank} and $Q_p$ is the particle (or normal-ordered)
charge. The total charge $Q$ is always conserved. For $\lambda $ just larger
than $\pi /2$, $Q_p=+1$ because of the presence of the positron and so the
vacuum charge $Q_0$ must now equal $-1$ to conserve charge$.$ As the
potential is increased further, $\lambda $ will reach $\pi .$ Here $E=-m$
which is the condition for supercriticality: the bound positron reaches the
continuum and becomes free. This is the well-known scenario of spontaneous
positron production first discussed \cite{zeld}\cite{grein2} over 25 years
ago. Note that at supercriticality $\lambda =\pi $, the even bound state
disappears and the first odd state appears.

\smallskip\ 

\noindent We can continue to increase $\lambda $ and count positrons: the
total number of positrons produced for a given $\lambda $ is the number of
times $E$ has crossed $E=0;$ that is 
\begin{equation}
Q_p=Int\,[\frac \lambda \pi +\frac 12]  \label{charge0}
\end{equation}

\noindent where $Int[x]$ denotes the integer part of $x.$ The more
interesting quantity for us is the number of supercritical positrons $Q_S$:
the number of states which have crossed $E=-m$. This is given by

\begin{equation}
Q_S=Int\,[\frac \lambda \pi ]  \label{super}
\end{equation}

\noindent Note that for any $\lambda $ there is at most one bound positron
state.

\subsection{Wide Well}

\noindent We consider the general case of $V>2m$ and then look in particular
at the case $ma>>1$ most closely corresponding to the Klein step. We must
find first the condition for supercriticality of the potential $V$, and then
the number of bound and supercritical positrons produced for a given $V.$
From Eq ( \ref{even}) we see that the ground state becomes supercritical
when $pa=\pi /2$ and therefore $V_1^c=m+\sqrt{m^2+\pi ^2/4a^2}. $ From Eq ( 
\ref{odd}) the first odd state becomes supercritical when $pa=\pi $ and $%
V_2^c=m+\sqrt{m^2+\pi ^2/a^2}.$ Clearly the supercritical potential
corresponding to the Nth positron is

\begin{equation}
V_N^c=m+\sqrt{m^2+N^2\pi ^2/4a^2}  \label{Nth}
\end{equation}

\noindent It follows from Eq (\ref{Nth}) that $V=2m$ is an accumulation
point of supercritical states as $ma\rightarrow \infty $. Furthermore it is
a threshold: a potential $V$ is subcritical if $V<2m$. It is not difficult
to show for a given $V>2m$ that the number of supercritical positrons is
given by

\begin{equation}
Q_S=Int[(2a/\pi )\sqrt{V^2-2mV}]  \label{super2}
\end{equation}

\noindent The corresponding value of the total positron charge $Q_p$ can be
shown u$\sin $g Eqs (\ref{even},\ref{odd}) to satisfy

\begin{equation}
Q_p-1\leq Int[(2a/\pi )\sqrt{V^2-m^2}]\leq Q_p  \label{charge}
\end{equation}

\noindent so for large $a$ we have the estimates

\begin{equation}
Q_p\sim (2a/\pi )\sqrt{V^2-m^2};\quad Q_S\sim (2a/\pi )\sqrt{V^2-2mV}
\label{est}
\end{equation}
$\qquad $

\noindent Now we can build up an overall picture of the wide square well $%
ma>>1$. When $V$ is turned on from zero in the vacuum state an enormous
number of bound states is produced. As $V$ crosses $m$ a very large number $%
Q_p$ of these states cross $E=0$ and become bound positrons. As $V$ crosses $%
2m$ a large number $Q_S$ of bound states become supercritical together. This
therefore gives rise to a positively charged current flowing from the well.
But in this case, unlike that of the Klein step, the charge in the well is
finite and therefore the particle emission process has a finite lifetime.
Nevertheless, for $ma$ large enough the transient positron current for a
wide barrier is approximately constant in time for a considerable time as we
shall see in the next section.\ 

\section{Emission Dynamics}

\noindent We now restrict ourselves to the case $V=2m+\Delta $ with $\Delta
<<m$. This is not necessary but it avoids having to calculate the dynamics
of positron emission while the potential is still increasing beyond the
critical value. We can assume all the positrons are produced almost
instantaneously as the potential passes through $V=2m.$ It also means that
the kinematics are non-relativistic. Hence for a sufficiently wide well so
that $\Delta a$ is large, $Q_S\sim 4\Delta a/\pi $. The well momentum of the
Nth supercritical positron is still given by Eq (\ref{quant}) $p_Na=N\pi /2$
which corresponds to an emitted positron energy $|\,E_N\,|=2m+\Delta -\sqrt{%
p_N^2+m^2}>m$. Note that the emitted energies have discrete values although
for $a$ large, they are closely spaced.

\smallskip\ 

\noindent The lifetime $\tau $ of the supercritical well is given by the
time for the slowest positron to get out of the well. The slowest positron
is the deepest lying state with $N=1$ and momentum $p_1=\pi /2a$. Hence $%
\tau \approx ma/p_1=2ma^2/\pi .$ So the lifetime is finite but scales as $%
a^2 $. But a large number of positrons will have escaped well before $\tau $%
. There are $Q_S$ supercritical positrons initially and their average
momentum $\overline{p}$ corresponds to $N=Q_S/2$; hence $\overline{p}=\Delta 
$ which is independent of $a$. Thus a transient current of positrons is
produced which is effectively constant in time for a long time of order $%
\overline{\tau }$ $=ma/\Delta $. We thus see that the square well (or
barrier) for $a$ sufficiently large behaves just like the Klein step: it
emits a seemingly constant current with a seemingly continuous energy
spectrum. But initially the current must build up from zero and eventually
must return to zero. So the well/barrier is a time-dependent physical entity
with a finite but long lifetime for emission of supercritical positrons or
electrons. In the Appendix we show that if we assume that the transient
current emitted is constant in time (which it is not), then it is possible
to obtain a relationship between this current and the transmission
coefficient just as for the Klein step.

\smallskip\ 

\noindent Note again that the transmission resonances of the
time-independent scattering problem coincide with the energies of particles
emitted by the well or barrier. It is therefore tempting to use the Pauli
principle to explain the connection. Following Hansen and Ravndal, we could
say that $R$ must be zero at the resonance energy because the electron state
is already filled by the emitted electron with that energy. But it is easy
to show that the reflection coefficient is zero for bosons as well as
fermions of that energy, and no Pauli principle can work in that case.
Furthermore emission ceases after time $\tau $ whereas $R=0$ for times $%
t>\tau $ . It follows that we must conclude that Klein tunnelling is a
physical phenomenon in its own right, independent of any emission process.
In our next section we consider further ways to investigate this tunnelling
theoretically and experimentally.

\section{Klein Tunnelling and the Coulomb Barrier}

\noindent It is clear from Eq (\ref{barr}) that while the reflection
coefficient $R$ cannot be $0$, neither is the transmission coefficient $T$
exponentially small for energies $E<V$, even though the scattering is
classically forbidden. The simplest way to understand this is to consider
the negative energy states under the potential barrier as corresponding to
physical particles which can carry energy in exactly the same way that
positrons are described by negative energy states which can carry energy. It
follows from Eq (\ref{Rs/Ts}) that $R_S$ and $T_S$ correspond to reflection
and transmission coefficients in transparent media with differing refractive
indices: thus $\kappa $ is nothing more than an effective fermionic
refractive index corresponding to the differing velocities of propagation by
particles in the presence and absence of the potential. On this basis,
tuning the momentum $p$ to obtain a transmission resonance for scattering
off a square barrier is nothing more than finding the frequency for which a
given slab of refractive material is tranparent. This is not a new idea. In
Jensen's words ''A potential hill of sufficient height acts as a Fabry-Perot
etalon for electrons, being completely transparent for some wavelengths,
partly or completely reflecting for others'' \cite{bak}.

\smallskip\ 

\noindent We can now look in more detail at Klein tunnelling: both in terms
of our model square well/barrier problem and at the analogous Coulomb
problem. The interesting region is where the potential is strong but
subcritical so that emission dynamics play no role and sensible time
independent scattering parameters can be defined. For electrons scattering
off the square barrier $V(x)=V$ we would thus require $V<$ $V_1^c=m+\sqrt{%
m^2+\pi ^2/4a^2}$ together with $V>2m$ so that positrons can propagate under
the barrier. For the corresponding square well $V(x)=-V$ there are negative
energy bound states $0>E>-m$ provided that $V>$ $\sqrt{m^2+\pi ^2/4a^2}$
[cf. Eq.(\ref{charge})]. So when the potential well is deep enough, it will
in fact bind positrons. Correspondingly, a high barrier will bind electrons.
It is thus not surprising that electrons can tunnel through the barrier for
strong subcritical potentials since they are attracted by those potential
barriers. Another way of seeing this phenomenon is by using the concept of
effective potential $V_{eff}(x)$ which is the potential which can be used in
a Schrodinger equation to simulate the properties of a relativistic wave
equation. For a potential $V(x)$ introduced as the time-component of a
four-vector into a relativistic wave equation (Klein-Gordon or Dirac), it is
easy to see that $2mV_{eff}(x)=2EV(x)-V^2(x).$ Hence as the energy $E$
changes sign, the effective potential can change from repulsive to
attractive.

\smallskip\ 

\noindent For the Coulomb barrier, Anchishkin \cite{anch} has already
suggested looking at scattering of $\pi ^{+}$ off heavy nuclei to see if
there was any experimental evidence for tunnelling: the Klein-Gordon
equation like the Dirac equation has negative energy solutions, so similar
arguments apply. His calculations show a 20\% enhancement in $\pi
^{+}-^{238}U$ scattering compared with non-relativistic expectations. The
analogous process for fermions would be positron-heavy nucleus scattering.
For a positron of initial energy $E$ incident on a heavy nucleus of charge $%
Z $ the classical turning point $r_c=Z\alpha /E.$ So it would be interesting
to measure the wave function at the origin for positron scattering off a
heavy nucleus compared with the wave function at the origin for electron
scattering off the nucleus at the same energy in order to demonstrate
tunnelling.

\smallskip\ 

\noindent Provided that $Z\alpha <1,E>0$ and normal Coulomb wave functions
should be accurate enough for the calculation of the ratio $\rho =\left|
\psi (0)\right| _{pos}^2/\left| \psi (0)\right| _{el}^2$ of wave functions
at the origin. The wave function at the origin for a Dirac particle is
singular but the ratio is finite. The exact continuum wave functions for a
Dirac particle in a Coulomb potential are discussed by Rose\cite{rose}. We
shall just write down his results: if the particles are non-relativistic then

\begin{equation}
\rho =e^{-2\pi Z\alpha E/p}  \label{nonrel}
\end{equation}
\noindent where p and $E$ are the particle momenta and energies and this is
ofcourse exponentially small as $p\rightarrow 0$. But if the particles are
relativistic 
\begin{equation}
\rho =fe^{-2\pi Z\alpha }  \label{rel}
\end{equation}
\noindent where $f$ is a ratio of complex gamma-functions and is
approximately unity for large $Z$. So $\rho \sim e^{-2\pi Z\alpha }\approx
10^{-3}$ for $Z\alpha \sim 1$which is not specially small although it still
decreases exponentially with $Z.$ It should be possible to measure $\rho $
by, for example, internal electron-positron pair production in
positron-nucleus scattering versus pair production in electron-nucleus
scattering. Alternatively if the positrons were longitudinally polarised any
resultant asymmetry in the scattering cross section would be due to weak
processes. These could only occur at a distance $r_w=1/M_Z$ which is
effectively at the origin for atomic systems. So the asymmetry for polarised
positron scattering could be compared with that for polarised electron
scattering. The ratios of the asymmetries would give $\rho $ directly.

\smallskip\ 

\noindent But while experiments for $Z\alpha <1$ should show tunnelling it
will not yet amount to Klein tunnelling. For that we require $Z$ large
enough so that bound positron states are present. This means that $Z$ must
be below its supercritical value $Z_c$ of around $170$ but large enough for
the $1s$ state to have $E<0$. The calculations of references \cite{zeld} and 
\cite{grein2} which depend on particular models of the nuclear charge
distribution give this region as $150<Z<Z_c$ which unfortunately will be
difficult to demonstrate experimentally. Nevertheless, the theory seems to
be clear: in this subcritical region positrons should no longer obey a
tunnelling relation which decreases exponentially with Z such as that of Eq.
(\ref{rel}). Instead the Coulomb barrier should become more transparent as $%
Z $ increases, at least for some energies. The work of Jensen and his
colleagues \cite{jens} shows that a transmission resonance (i.e. maximal
transmission) for positron scattering on a Coulomb potential may well occur
at $Z=Z_c$ although the onset of supercriticality implies that time
independent scattering quantities will then no longer be well-defined.
Further detailed calculations are needed to clarify the situation for
positron scattering off nuclei with $Z$ just below $Z_c$. It is also
important to see if the square well/barrier relationship we illustrate above
that transmission resonances at barriers occur when the corresponding well
becomes supercritical [cf. Eqs (\ref{quant}) and (\ref{even}] can be
generalised to arbitrary potentials.

\section{Conclusions}

\noindent We hope that this discussion has demonstrated that the Klein step
is pathological and therefore a misleading guide to the underlying physics.
It represents a limit in which time-dependent emission processes become
time-independent and therefore a relationship between the emitted current
and the transmission coefficient exists, as we show in the Appendix. In
general no such relationship exists. The underlying physics discovered by
Klein in his solution of the Dirac equation is not particle emission but
tunnelling by means of the negative energy states which are characteristic
of relativistic wave equations, similar to interband tunnelling in
semiconductors \cite{semi}. It is time to finally bring this 70 year old
puzzle to a conclusion.

\smallskip\ 

\noindent We would like to thank A. Anselm, G Barton, J D Bjorken, L B Okun,
R Laughlin, G E Volovik and D Waxman for advice and help.

\section{Appendix: The calculation of the vacuum current in the presence of
a Klein step.}

\noindent Consider the potential step $V(x)=V,x>0;\,V(x)=0,x<0$ for $V>2m$
of section I. We will show that the expectation value of the current in the
vacuum state in the presence of the step is non-zero. The derivation hinges
on a careful definition of the vacuum state.

\subsection{The normal modes in the presence of the Klein step.}

\noindent A properly-normalised positive energy solution to the Dirac
equation (\ref{dir}) can be written

\begin{equation}
\sqrt{%
%TCIMACRO{\dfrac{E+m}{2k} }
%BeginExpansion
{\displaystyle {E+m \over 2k}}
%EndExpansion
}\left( 
\begin{array}{c}
i \\ 
%TCIMACRO{\dfrac k{E+m}}
%BeginExpansion
{\displaystyle {k \over E+m}}
%EndExpansion
\end{array}
\right) e^{ikx}  \label{pos}
\end{equation}

\noindent Scattering is usually described by a solution describing a wave
incident (say from the left) plus a reflected wave ( from the right) plus a
transmitted wave (to the right). It is convenient here to use waves of
different form either describing a wave (subscript $L)$ incident from the
left with no reflected wave or describing a wave (subscript $R$) incident
from the right with no reflected wave. Positive and negative energy
wavefunctions will be denoted by $u$ and $v$ respectively. It is clear that
the nontrivial result we are seeking arises from the overlap of the negative
energy continuum $E<V-m$ on the right with the positive energy continuum $%
E>m $ on the left. We are thus concerned with wavefunctions with energies in
the range $m<E<V-m.$ The expressions for $u_L,u_R$ in this energy range are
given below.

\ 

\begin{equation}
\begin{array}{c}
u_L(E,x)=%
%TCIMACRO{\dfrac{\sqrt{2\kappa }}{\kappa +1} }
%BeginExpansion
{\displaystyle {\sqrt{2\kappa } \over \kappa +1}}
%EndExpansion
\sqrt{%
%TCIMACRO{\dfrac{E+m}k }
%BeginExpansion
{\displaystyle {E+m \over k}}
%EndExpansion
}\left( 
\begin{array}{c}
i \\ 
%TCIMACRO{\dfrac k{E+m}}
%BeginExpansion
{\displaystyle {k \over E+m}}
%EndExpansion
\end{array}
\right) e^{ikx}\theta (-x)+ \\ 
\\ 
\left\{ 
%TCIMACRO{\dfrac{\kappa -1}{\kappa +1} }
%BeginExpansion
{\displaystyle {\kappa -1 \over \kappa +1}}
%EndExpansion
\sqrt{%
%TCIMACRO{\dfrac{V-E-m}{2\left| p\right| } }
%BeginExpansion
{\displaystyle {V-E-m \over 2\left| p\right| }}
%EndExpansion
}\left( 
\begin{array}{c}
i \\ 
%TCIMACRO{\dfrac{\left| p\right| }{E+m-V}}
%BeginExpansion
{\displaystyle {\left| p\right|  \over E+m-V}}
%EndExpansion
\end{array}
\right) e^{i\left| p\right| x}+\sqrt{%
%TCIMACRO{\dfrac{V-E-m}{2\left| p\right| } }
%BeginExpansion
{\displaystyle {V-E-m \over 2\left| p\right| }}
%EndExpansion
}\left( 
\begin{array}{c}
i \\ 
%TCIMACRO{\dfrac{-\left| p\right| }{E+m-V}}
%BeginExpansion
{\displaystyle {-\left| p\right|  \over E+m-V}}
%EndExpansion
\end{array}
\right) e^{-i\left| p\right| x}\right\} \theta (x)
\end{array}
\label{w1}
\end{equation}

\begin{equation}
\begin{array}{c}
u_R(E,x)=\left\{ 
%TCIMACRO{\dfrac{1-\kappa }{1+\kappa } }
%BeginExpansion
{\displaystyle {1-\kappa  \over 1+\kappa }}
%EndExpansion
\sqrt{%
%TCIMACRO{\dfrac{E+m}{2k} }
%BeginExpansion
{\displaystyle {E+m \over 2k}}
%EndExpansion
}\left( 
\begin{array}{c}
i \\ 
%TCIMACRO{\dfrac k{E+m}}
%BeginExpansion
{\displaystyle {k \over E+m}}
%EndExpansion
\end{array}
\right) e^{ikx}+\sqrt{%
%TCIMACRO{\dfrac{E+m}{2k} }
%BeginExpansion
{\displaystyle {E+m \over 2k}}
%EndExpansion
}\left( 
\begin{array}{c}
i \\ 
%TCIMACRO{\dfrac{-k}{E+m}}
%BeginExpansion
{\displaystyle {-k \over E+m}}
%EndExpansion
\end{array}
\right) e^{-ikx}\right\} \theta (-x)+ \\ 
\\ 
+%
%TCIMACRO{\dfrac{\sqrt{2\kappa }}{\kappa +1} }
%BeginExpansion
{\displaystyle {\sqrt{2\kappa } \over \kappa +1}}
%EndExpansion
\left( 
\begin{array}{c}
i \\ 
%TCIMACRO{\dfrac{\left| p\right| }{E+m-V}}
%BeginExpansion
{\displaystyle {\left| p\right|  \over E+m-V}}
%EndExpansion
\end{array}
\right) e^{i\left| p\right| x}\theta (x)
\end{array}
\label{w2}
\end{equation}

\noindent We write $\left| p\right| $ rather than $p$ in these equations
since as Pauli noted, the group velocity is negative for $x>0$. For $V-E=%
\sqrt{p^2+m^2}$ we have $dE/dp=-\left| p\right| \,/\sqrt{p^2+m^2}.$ This
gives the expressions above.

\smallskip\ 

\noindent We need to evaluate the currents corresponding to the solutions of
Eqs (\ref{w1},\ref{w2}). According to our conventions $\alpha _x\equiv
\gamma _0\gamma _x=-\sigma _y$ so 
\begin{equation}
j_L\equiv -u_L^{\dagger }(E,x)\sigma _yu_L(E,x)=\frac{2\kappa }{\pi (\kappa
+1)^2}  \label{c1}
\end{equation}
\begin{equation}
j_R\equiv -u_R^{\dagger }(E,x)\sigma _yu_R(E,x)=-\frac{2\kappa }{\pi (\kappa
+1)^2}  \label{c2}
\end{equation}

\subsection{The definition of the vacuum and the vacuum expectation value of
the current.}

We expand $\psi $ in terms of creation and annihilation operators terms in
terms of our left- and right-travelling solutions: 
\begin{equation}
\begin{array}{c}
\psi (x,t)=\int dE\{a_L(E)u_L(E,x)e^{-iEt}+a_R(E)u_R(E,x)e^{-iEt}+ \\ 
\\ 
+b_L^{\dagger }(E)v_L(E,x)e^{iEt}+b_R^{\dagger }(E)v_R(E,x)e^{iEt}\}
\end{array}
\label{exp}
\end{equation}

\noindent with $\psi ^{\dagger }$ given by the Hermitean conjugate
expansion.We must now determine the appropriate vacuum state in the presence
of the step. States described by wavefunctions $u_L(E,x)$ and $v_L(E,x)$
correspond to (positive energy) electrons and positrons respectively coming
from the left. Hence with respect to an observer to the left (of the step)
such states should be absent from the vacuum state, so

\begin{equation}
a_L(E)\left| 0\right\rangle =0,\,b_L(E)\left| 0\right\rangle =0  \label{a1}
\end{equation}

\noindent Wavefunctions $u_R(E,x)$ for $E>m+V$ describe for an observer to
the right, electrons incident from the right. These are not present in the
vacuum state hence 
\begin{equation}
a_R(E)\left| 0\right\rangle =0\text{ for }E>m+V  \label{a2}
\end{equation}

\noindent Wavefunctions $v_R(E,x)$ describe, again with respect to an
observer to the right, positrons incident from the right; again 
\begin{equation}
b_R(E)\left| 0\right\rangle =0\text{ }  \label{b2}
\end{equation}

\noindent The wavefunctions that play the crucial role in the Klein problem
belong to the set $u_R(E,x)$ for $m<E<V-m.$ For an observer to the right
these states are positive energy positrons and hence they should be filled
in the vacuum state, i.e.

\begin{equation}
a_R^{\dagger }(E)a_R(E^{\prime })\left| 0\right\rangle =\delta (E-E^{\prime
})\left| 0\right\rangle \text{ , }m<E<V-m  \label{vac}
\end{equation}

\noindent Having specified the vacuum the next and final step is the
calculation of the vacuum expectation value \ of the current: 
\begin{equation}
\left\langle 0\right| j\left| 0\right\rangle =\frac 12\left( -\left\langle
0\right| \psi ^{\dagger }\sigma _y\psi \left| 0\right\rangle +\left\langle
0\right| \psi \sigma _y\psi ^{\dagger }\left| 0\right\rangle \right)
\label{vev}
\end{equation}

\noindent Substituting (\ref{exp}) in (\ref{vev}) and noticing that all
terms involving $v_L$ and $v_R$ can be dropped since the corresponding
energies lie outside the interesting range $m<E<V-m$ we end up with 
\begin{equation}
\begin{array}{c}
\left\langle 0\right| j\left| 0\right\rangle =-\frac 12\int dEdE^{\prime
}\{\left\langle 0\right| a_L^{\dagger }(E)a_L(E^{\prime })\left|
0\right\rangle u_L^{\dagger }(E,x)\sigma _yu_L(E^{\prime },x)+ \\ 
\\ 
+\left\langle 0\right| a_L(E)a_L^{\dagger }(E^{\prime })\left|
0\right\rangle u_L^{\dagger }(E^{\prime },x)\sigma _yu_L(E,x)-\left\langle
0\right| a_R^{\dagger }(E)a_R(E^{\prime })\left| 0\right\rangle u_R^{\dagger
}(E,x)\sigma _yu_R(E^{\prime },x)+ \\ 
\\ 
+\left\langle 0\right| a_R(E)a_R^{\dagger }(E^{\prime })\left|
0\right\rangle u_R^{\dagger }(E^{\prime },x)\sigma _yu_R(E,x)\}
\end{array}
\label{sum}
\end{equation}

\smallskip\ 

\noindent The first term in (\ref{sum}) vanishes due to (\ref{a1}). The
second term becomes

$u_L^{\dagger }(E^{\prime },x)\sigma _yu_L(E,x)\delta (E-E^{\prime })$ if we
use the anticommutation relations and (\ref{a1}). The third term yields $%
-u_R^{\dagger }(E,x)\sigma _yu_R(E,x)\delta (E-E^{\prime })$ using (\ref{vac}%
) and the fourth term vanishes using the anticommutation relations (i.e. the
exclusion principle; the state $\left| 0\right\rangle $ already contains an
electron in the state $u_R$ hence we get zero when we operate on it with $%
a_R^{\dagger }$). One energy integration is performed immediately using the $%
\delta $ function. The final result is

\begin{equation}
\left\langle 0\right| j\left| 0\right\rangle =\frac 12\int
dE(-j_L+j_R)=-\int dE\frac{4\kappa (E)}{(\kappa (E)+1)^2}\equiv -\int
dET_S(E)  \label{res}
\end{equation}

\noindent where the energy integration is over the Klein range. This is the
result first obtained by Hansen and Ravndal \cite{hans} linking the pair
production current with the transmission coefficient. But the fact that
these quantities are linked does not represent a causal relationship between
them. This is clear if we attempt to repeat this analysis for a square
barrier for $V>2m$ and $ma>>1$. It is easily shown by the same method that

\begin{equation}
\left\langle 0\right| j\left| 0\right\rangle =-\int dE\frac{8\kappa ^2}{%
(1-\kappa ^2)^2+8\kappa ^2}\equiv -\int dET_\infty (E)  \label{cur}
\end{equation}

\noindent provided that averages over phase angles are understood, where now
the current is measured by an observer situated at a point $x>a.$ The result
of Eq. (\ref{cur}) of course can only be true in the approximation where the
transient current is taken to be independent of time.

\smallskip\ 

\noindent Note also that whereas for a Klein step we describe the particle
emission in terms of electron-positron pair production while we say that a
supercritical well will spontaneously emit positrons. These terms are
relative: an observer outside a supercritical well would see a positron
current going to the right while an observer measuring the same phenomenon
but inside the well would talk of an electron current going to the left.

\end{document}